\documentclass[
reprint,
 amsmath,amssymb,
 aps,
prl,
]{revtex4-2}

\usepackage{graphicx}
\usepackage{dcolumn}
\usepackage{bm}
\usepackage{graphicx}

\usepackage{xcolor}

\begin{document}

\preprint{APS/123-QED}

\title{Measurement of the Quantum Capacitance Between Two Metallic Electrodes}

\author{T. de Ara}
\altaffiliation{Present address: Institute of Physics, École Polytechnique Fédérale de Lausanne (EPFL), CH-1015 Lausanne, Switzerland}
\author{B. Olivera}
\author{C. Sabater}
\author{C. Untiedt}
 \email{untiedt@ua.es}
\affiliation{Departamento de F\'\i sica 
 and Instituto Universitario de Materiales de Alicante (IUMA), Facultad de Ciencias. Universidad de Alicante, Campus de San Vicente del Raspeig, E-03690 Alicante, Spain.
}

\date{\today}

\begin{abstract}
Two factors contribute to the electrical capacitance between two electrodes: a classical contribution, stemming from the electric field, and a quantum contribution, governed by the Pauli exclusion principle, which increases the difficulty of adding charge to the electrodes.
In metals, the high electronic Density of States (DOS) at the Fermi energy allows the quantum contribution to be neglected, and a classical description of the electrical capacitance between two metallic electrodes is normally used. 
Here, we study the evolution of the capacitance as two metallic electrodes (Pt or Au) are approached to the limit when quantum corrections are needed, before contact formation. 
At small distances, we observe that the classical increase in capacitance turns into saturation as the electrodes are approached, reaching the quantum capacitance limit. Finally, a capacitance leakage due to quantum tunneling is observed. 
Since the quantum capacitance depends on the electronic DOS on the surface of the electrodes, we use it to probe the DOS change induced by molecular adsorption (Toluene) on the metallic surface.

\end{abstract}

\maketitle


Capacitance quantifies the accumulation of charges at the surfaces of two electrodes when a voltage difference is applied. Classically, in vacuum, capacitance depends only on the flux of the electric field that is developed between the electrodes, and therefore on their area, geometry, and separation.
Another contribution, often overlooked in the context of metals, comes from the quantum character when the Pauli exclusion principle limits charge accumulation on the electrode surface, known as quantum capacitance \cite{Mead61,Luryi88}.
This quantum contribution acts as a capacitor in series with the classical capacitance and is defined as the variation of electrical charge $q$ with respect to the variation of electrochemical potential $C_q=\frac{dq}{d\mu}=e^2\rho$, where $\rho$ is the Density of States (DOS) at the Fermi energy of the electrodes and $e$ is the electron charge.
While quantum capacitance has been proven to be relevant in low DOS systems, such as a 2D electronic system in a semiconductor surface \cite{Ando82,Luryi88,John04} or in 2D materials such as graphene \cite{Xia09,Miskovic010}, it is generally assumed to be negligible in metals.

Christen and B\"uttiker \cite{Buttiker96, Buttiker93} studied theoretically a mesoscopic capacitor with tunneling between its plates (leakage) with transmission probability $T$, showing that the capacitance ($C_{\rm T}$) would be proportional to the reflection probability $R=1-T$ of the electrons at the plates, and therefore, can be expressed as:
\begin{equation}\label{Chr-But}
    C_{\rm T}=(1-T)\left( C_0+\frac{1}{C_{\rm geo}^{-1}+C_{\rm \mu_1}^{-1}+C_{\rm \mu_2}^{-1}}
    \right)
\end{equation}
where, for experimental reasons, we have added the possibility of having a parallel capacitor $C_0$ that accounts for extra stray capacitance, $C_{\rm geo}$ is the classical or geometrical capacitance, and $C_{\rm \mu_x}$ {\{x=1,2\} } are the quantum capacitances of the two electrodes. 
For the case of a single transmission channel in the tunneling regime, the transmission probability would be written as:  
\begin{equation}\label{exp_dec}
T=e^{-2\sqrt{2m_{\rm e}\phi}\; d/\hbar}
\end{equation}
where $\phi$ is the work function of the electrodes and $d$ their distance. 

The role of this effect for metals has been explored for small objects in Scanning Tunneling Microscope (STM) studies. Theoretically, it was studied for the case of a 5-atom tip against a flat substrate using {\it ab initio} methods \cite{Wang98}, and experimentally for small clusters (1-10~nm diameter), using Coulomb-blockade characteristics to infer the capacitance in between the clusters and an STM tip \cite{Hou01}. 
In these studies, the dependence of the capacitance with the distance followed the behavior predicted by the previous expression with capacitance values below $aF$.  

Here, we study experimentally, within the framework of the Christen-B\"uttiker model, the capacitance of two macroscopic metallic electrodes being approached until the tunneling regime, with the aim of quantifying the relevance of the quantum contribution to the capacitance. As we will show below, this contribution becomes relevant for inter-electrode distances above tens of nanometers. The large electrode area involved in our experiments enhances our resolution and the saturation capacity to the order of $fF$, allowing us to overcome some of the limitations of previous studies\cite{Hou01}.

\begin{figure*}[ht] 
\includegraphics[width=1
\textwidth]{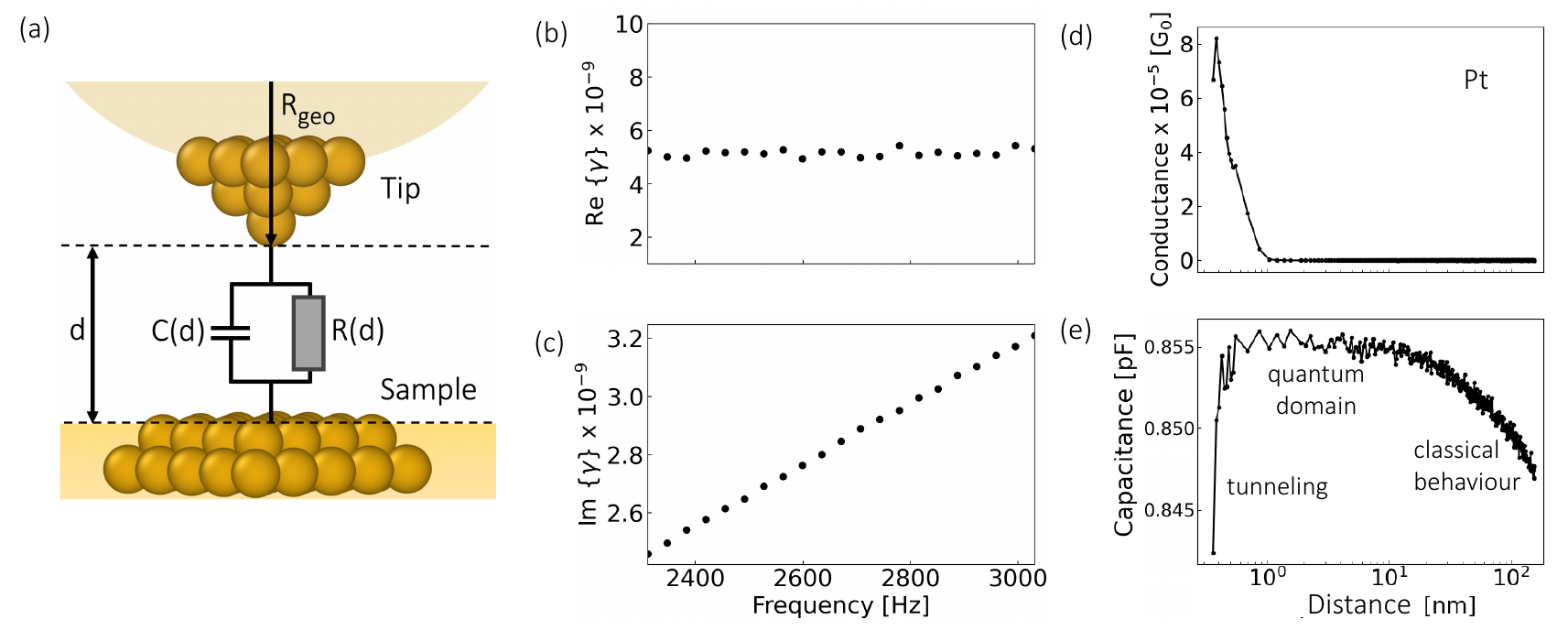}
\caption{\label{fig1} (a) Schematic illustration of the measurement of conductance and capacitance between two crossed metallic wires as a function of their separation distance. The equivalent circuit consists of a resistance and a capacitor connected in parallel between both electrodes. Panels (b) and (c) show an example of a frequency sweep for measuring the real and imaginary part of the admittance of the junction. Panels (d) and (e) display the experimental measurements of Pt wires conductance and capacitance at 77~K. In panel (e), we also label the three observed regions as tunneling, quantum, and classical domains.}
\end{figure*}

We have measured the variations of the capacitance with respect to the distance using a modified low-temperature STM on two wires of the same material, crossed against each other. The electrode wires were cylindrical, 0.5 mm in diameter, and of high purity 99.999~$\%$ Pt or Au.
As quantum capacitance is proportional to the surface DOS, the wire surfaces were carefully cleaned by exposing the metal to plasma.
The STM used for our experiments is a homemade STM based on the design by \cite{pan1999he} with a 20~pm stability and a travel range of approximately 0.5~$\mu$m in the $Z$ direction, which was used to vary the inter-electrode distance in our experiment. The STM was mounted on a He cryostat under cryogenic vacuum.   

For a complete electrical impedance characterization of our system, we performed a 4-probe AC measurement using two synchronized Signal Recovery 7265DSP Lock-in amplifiers. The 4-probe measurement configuration allows us to reduce to the minimum the effect of the impedance of the equipment in our measurement.

In our experiments, we measured the impedance of the system at different distances as the two electrodes are moving apart. For each distance point, the impedance was recorded over a range of frequencies. In panels (b) and (c) of Fig.\ref{fig1}, a typical impedance measurement at a given distance is shown. 
This measurement offers a value of both resistance and capacitance, by using the real and imaginary parts, respectively. 
An average value across the frequency sweep of the real component provides the inverse of the resistance, i.e., the conductance (see panel (d)), while the slope of the imaginary component yields the capacitance, obtained from a least square fitting (as displayed in panel (e)).

A typical result of our measurements is shown in Fig. \ref{fig1}, for Pt at 77 K under cryogenic vacuum, where the evolution of capacitance and conductance is recorded.
In the conductance, we observe an exponential decay as the distance increases, consistent with the tunneling behaviour described by Eq.\ref{exp_dec}.
Meanwhile, the capacitance (bottom panel) shows first an initial increase, followed by a sort of plateau, and finally a decrease of its value at large distances. It should be stressed here that, to make these three different regimens clear, the distance is plotted on a logarithmic scale. 

The three regimes observed in the capacitance can be understood within the framework of Christen and B\"uttiker's theory, and hence, to the different terms in Eq.\ref{Chr-But}.  
At short distances, the increase in capacitance is related to the term $(1-T)$ and, therefore, to the decrease in conductance. This can be clearly seen in the experiments as there is a one-to-one correlation between the tunneling decrease of conductance and its correlative increase of capacitance. 
On another hand, at large distances, the decrease of capacitance has its origin in the geometrical or classical capacitance. In our case, we have approached the geometry of our system as two spheres facing each other\cite{Crowley08}:
\begin{equation}\label{C_geo}
    C_{\rm geo}(d)=4\pi \epsilon R \left( 1+{1\over 2} \log \left(1+ {R \over d} \right) \right)
\end{equation}
where $R$ is the wire radius. This model is a rough approximation of the actual geometry, as it does not capture the geometry of the tip, the atomic structure, or the electronic interactions. However, it provides an estimation of the wire radius at large distances. In fact, we have observed a good agreement between our fits to this model for different samples to the actual radius of the used wires.

In previous attempts to understand the capacitance for the case of STM experiments \cite{Kurokawa98}, it was suggested that the shape of the capacitative curve was only related to geometrical factors, and therefore to the shape of the tip. In this case, then, Eq.\ref{Chr-But} would be reduced to $C_{\rm classic}~=~(1-T) (C_0 +C_{\rm geo}(d))$. 

\begin{figure}[ht]
\includegraphics[width=0.49\textwidth]{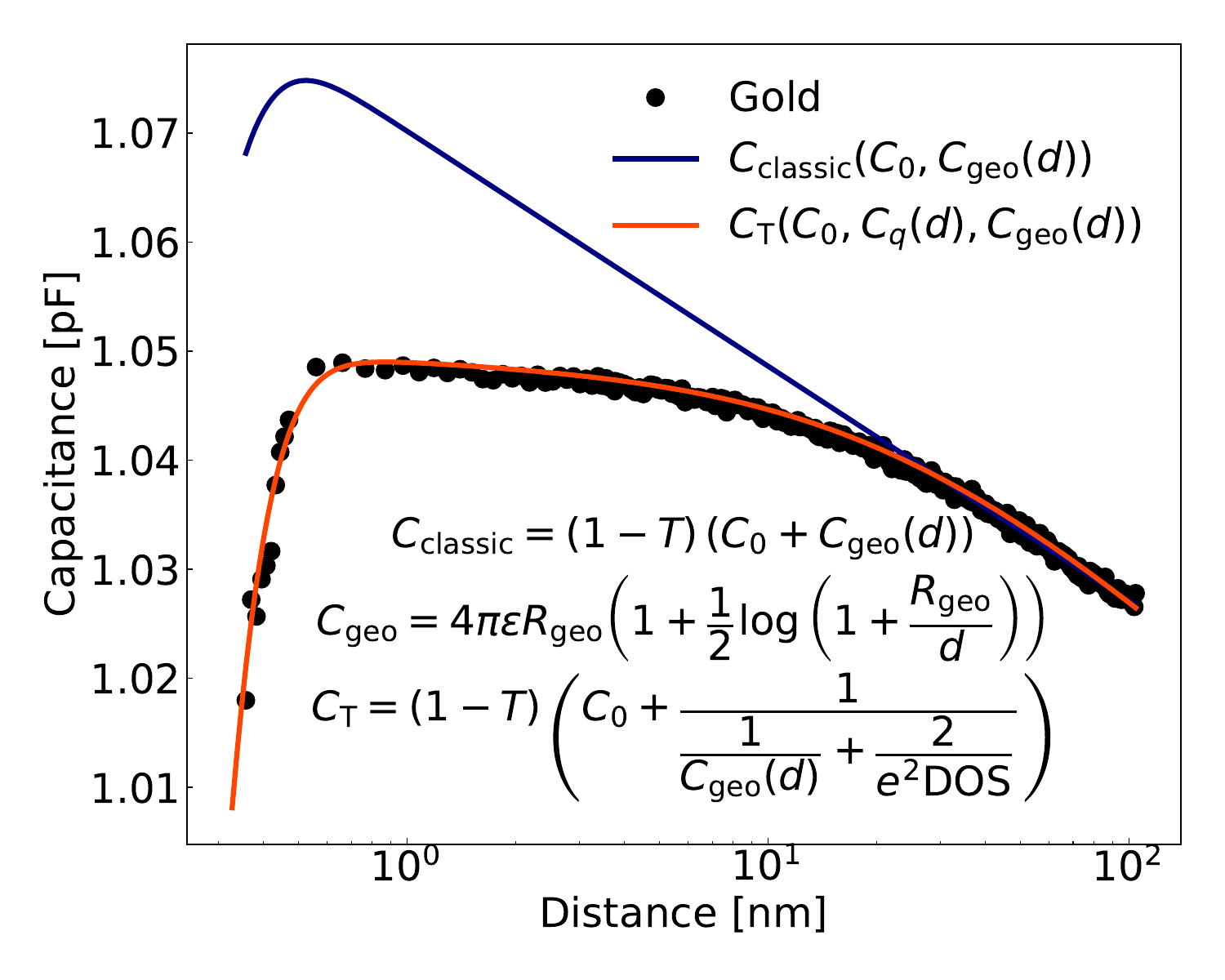}
\caption{ \label{fig2} Fitting of the capacitance curve for two gold electrodes in cryogenic vacuum at 4.2~K.
For the blue curve, we neglect the quantum capacitance contribution. The fitting for large distances gives the actual radius of our wire samples $R_{\rm geo}$~=~0.25~mm and a stray capacitance parameter $C_0~=~0.94$~pF. When we include the quantum capacitance term, the red-orange curve is obtained, which fits perfectly our experimental data for the case of an electronic DOS per unit area, $\rho_A$~=~$1.15\cdot10^{18}$~eV$^{-1}$m$^{-2}$.}
\end{figure}

As we can determine $T$ from the conductance curve and $C_{\rm geo}(d)$ from Eq.\ref{C_geo} for long distances, we can fit our experimental data with only $C_0$ as a free parameter. This fit is shown in Fig.\ref{fig2} (blue curve). However, the fit fails for distances below 20 nm. 
In order to understand the results of the work of Kurokawa \textit{et al.}\cite{Kurokawa98}, we pushed this model further to try fitting our data. The only way we could make it work was by artificially adding $d_0$~=~25 nm to the $C_{\rm geo}$. Although a remarkably good fit of the data is obtained, there is no physical justification for introducing this extra adjustment parameter.
One key difference between their work and ours is that in our case, we have a reliable reference for $d=0$ from the tunneling current, and therefore, this suggests the incompleteness of this previous model based only on geometrical factors. 

All of the above suggest the necessity of including the quantum capacitance contribution in the model to understand our data. 
We should remember that quantum capacitance can be written as $C_q=~e^2 \int_{-\infty} ^\infty DOS(\epsilon)f(\epsilon)d\epsilon$ where $f(\epsilon)$ is the Fermi distribution and $DOS(\epsilon)~=~A\cdot~\rho_A(\epsilon)$ is the total density of states integrated in the area $A$ where charge is accumulated.
For including this new additional parameter in Eq.\ref{Chr-But}, we consider the contribution of the quantum capacitance to be the same for both electrodes $C_{\mu_1}=C_{\mu_2}$, as their geometry and electronic DOS can be considered equal. 
On the other hand, the dependence of the effective area of quantum capacitance with distance can be considered to be the same as the effective area that leads to the expression in Eq.\ref{C_geo}, leaving a single new fitting parameter for our data, the electronic DOS per unit area, $\rho_A$.

The fit to our experimental data points using the contribution of quantum capacitance is shown by the red curve in Fig.\ref{fig2}, where we obtain a DOS per unit area 
$\rho_A~=~1.15 \cdot 10^{18}~\rm {eV}^{-1}m^{-2}$. 
This value is consistent with an estimation based on bulk of $\rho_A~=~8.9\cdot~10^{17}~\rm eV^{-1}$m$^{-2} \, (\rho_{\rm atom}=0.24~\rm eV^{-1}$ per atom~\cite{roth2017electronic} and atomic area $a~$=~0.268~$\rm nm^2$). The agreement between the fitted curve and the experimental data is remarkable, highlighting the importance of the quantum contribution in our measurements.

Nevertheless, to further demonstrate the validity of our model, we considered the modification of the quantum capacitance without affecting the geometrical part.
As quantum capacitance reflects the DOS around the Fermi energy, any change or contamination at the surface of the electrodes should lead to changes in the measured capacitance. This fact can be used to prove the role of the quantum capacitance in our system.

For this purpose, we have used Toluene, which, as shown in previous experiments \cite{DeAra22, Martinez23}, leaves a thin layer (probably a monolayer) on the gold surface. 
In our experiment, to avoid any change in the geometrical characteristics of our system, after characterizing the capacitative characteristics of two Au electrodes from room temperature down to Helium temperature, we covered the surface of the electrodes with Toluene and let it evaporate. The resulting thin layer of these molecules at the gold surface should be enough to modify the surface electronic DOS of the electrodes.  

\begin{figure}[ht]
\includegraphics[width=0.45\textwidth]{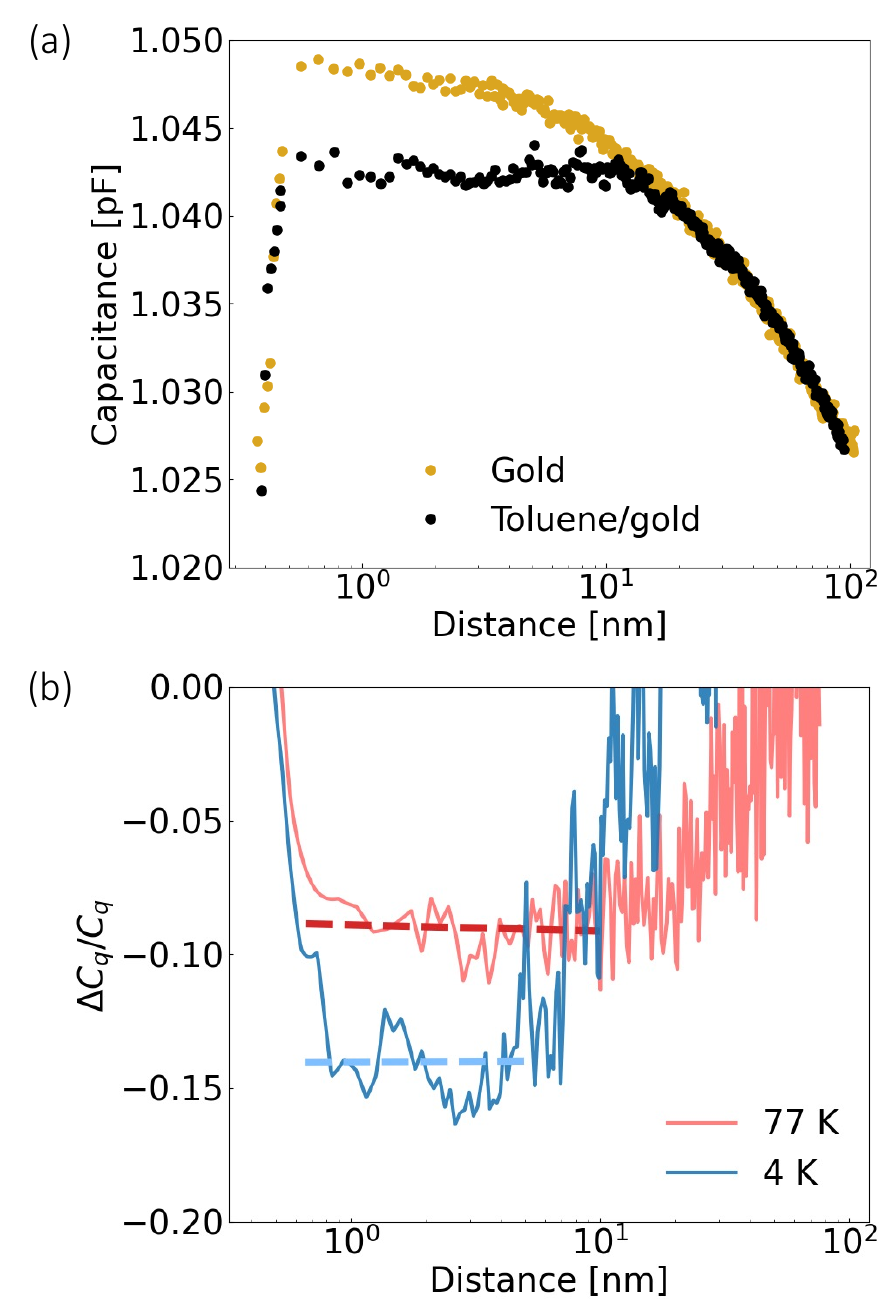}
\caption{\label{fig4} The upper panel shows the capacitance curves for clean Au electrodes and the same but Toluene-covered, measured at 4~K. In the lower panel, the difference in quantum capacitance is numerically extracted, showing a reduction of about 10\%. In panel (b), the results for 77~K are also shown in light red. Notice that for long distances, the dominant geometrical capacitance makes it difficult to appreciate changes in the quantum contribution.}
\end{figure}
\begin{acknowledgments}

In the upper panel of Fig.\ref{fig4}, we compare the curves obtained for clean (dark yellow) and Toluene-covered (black) samples at 4.2~K. A clear difference appears for short distances, below $\sim 20$~nm, where the quantum contribution is prominent. By numerically subtracting the two curves, we obtain the difference in quantum capacitance, as shown in the lower panel of Fig.\ref{fig4}, revealing a reduction of the quantum capacitance in the covered sample that can be quantified. Our results show a $\sim$13\% decrease in the quantum capacitance when a thin Toluene layer is adsorbed on the gold surface, coming from the modification in its electronic DOS.  
One should notice that the difference is hardly noticeable at distances above 10~nm because the geometrical contribution is dominant. This highlights the importance of small distances to get information on the quantum capacitance. 
We have also noticed that the difference in quantum capacitance increases with decreasing temperature, going from 9~\% at 77~K to 13~\% at 4.2~K (see panel b). This trend could be related to the temperature dependence of the electronic DOS (DOS~$\sim$~$\sqrt{E}$ with E~=~$k_B T$, where $k_B$ is the Boltzmann constant) and should be further investigated, although this is not within the scope of the present work.  

This result with the Toluene-covered sample should also explain the variability observed in different experiments for the value of the quantum capacitance parameter. The strong dependence on the cleanliness of our samples in a large area of 16~$\mu m^2$ at $d$~=~10 nm (with $A_{\mathrm{eff}} \approx 2\pi R_{\mathrm{geo}} d$ in the sphere-plane approximation) around the central part of our capacitor can substantially modify the shape of our curves at short distances, as we have observed.

In summary, we have demonstrated the relevance of the contribution of quantum capacitance in the case of metals to capacitance when the separation between the electrodes is even several tens of nanometers. The high sensitivity of this effect could be important for understanding the behavior of electronic devices. It can also be used to characterize variations of the surface electronic DOS coming from different sources, for example, in the case of having, as we have demonstrated, small amounts of contaminants at the surface, opening the way for a new method for molecular sensing.    

We acknowledge useful discussions with J. Fern\'{a}ndez-Rossier and D. Jacob. This work forms part of the Advanced Materials program and was supported by MCIN with funding from European Union NextGenerationEU (PRTR-C17.I1) and by Generalitat Valenciana (MFA/2022/045). The authors acknowledge financial support from the Spanish Government through   PID2022-141712NB-C22 and by the Generalitat Valenciana through PROMETEO/2021/017 and CIDEXG/2022/45. We also acknowledge funding from MICIU/AEI/10.13039/501100011033 and the European Regional Development Fund (ERDF/EU) under project PID2023-146660OB-I00.
\end{acknowledgments}

\bibliography{capacitance}

\providecommand{\noopsort}[1]{}\providecommand{\singleletter}[1]{#1}%
\begin{thebibliography}{16}%
\makeatletter
\providecommand \@ifxundefined [1]{%
 \@ifx{#1\undefined}
}%
\providecommand \@ifnum [1]{%
 \ifnum #1\expandafter \@firstoftwo
 \else \expandafter \@secondoftwo
 \fi
}%
\providecommand \@ifx [1]{%
 \ifx #1\expandafter \@firstoftwo
 \else \expandafter \@secondoftwo
 \fi
}%
\providecommand \natexlab [1]{#1}%
\providecommand \enquote  [1]{``#1''}%
\providecommand \bibnamefont  [1]{#1}%
\providecommand \bibfnamefont [1]{#1}%
\providecommand \citenamefont [1]{#1}%
\providecommand \href@noop [0]{\@secondoftwo}%
\providecommand \href [0]{\begingroup \@sanitize@url \@href}%
\providecommand \@href[1]{\@@startlink{#1}\@@href}%
\providecommand \@@href[1]{\endgroup#1\@@endlink}%
\providecommand \@sanitize@url [0]{\catcode `\\12\catcode `\$12\catcode `\&12\catcode `\#12\catcode `\^12\catcode `\_12\catcode `\%12\relax}%
\providecommand \@@startlink[1]{}%
\providecommand \@@endlink[0]{}%
\providecommand \url  [0]{\begingroup\@sanitize@url \@url }%
\providecommand \@url [1]{\endgroup\@href {#1}{\urlprefix }}%
\providecommand \urlprefix  [0]{URL }%
\providecommand \Eprint [0]{\href }%
\providecommand \doibase [0]{https://doi.org/}%
\providecommand \selectlanguage [0]{\@gobble}%
\providecommand \bibinfo  [0]{\@secondoftwo}%
\providecommand \bibfield  [0]{\@secondoftwo}%
\providecommand \translation [1]{[#1]}%
\providecommand \BibitemOpen [0]{}%
\providecommand \bibitemStop [0]{}%
\providecommand \bibitemNoStop [0]{.\EOS\space}%
\providecommand \EOS [0]{\spacefactor3000\relax}%
\providecommand \BibitemShut  [1]{\csname bibitem#1\endcsname}%
\let\auto@bib@innerbib\@empty
\bibitem [{\citenamefont {Mead}(1961)}]{Mead61}%
  \BibitemOpen
  \bibfield  {author} {\bibinfo {author} {\bibfnamefont {C.~A.}\ \bibnamefont {Mead}},\ }\bibfield  {title} {\bibinfo {title} {Anomalous capacitance of thin dielectric structures},\ }\href {https://doi.org/10.1103/PhysRevLett.6.545} {\bibfield  {journal} {\bibinfo  {journal} {Phys. Rev. Lett.}\ }\textbf {\bibinfo {volume} {6}},\ \bibinfo {pages} {545} (\bibinfo {year} {1961})}\BibitemShut {NoStop}%
\bibitem [{\citenamefont {Luryi}(1988)}]{Luryi88}%
  \BibitemOpen
  \bibfield  {author} {\bibinfo {author} {\bibfnamefont {S.}~\bibnamefont {Luryi}},\ }\bibfield  {title} {\bibinfo {title} {{Quantum capacitance devices}},\ }\href {https://doi.org/10.1063/1.99649} {\bibfield  {journal} {\bibinfo  {journal} {Applied Physics Letters}\ }\textbf {\bibinfo {volume} {52}},\ \bibinfo {pages} {501} (\bibinfo {year} {1988})}\BibitemShut {NoStop}%
\bibitem [{\citenamefont {Ando}\ \emph {et~al.}(1982)\citenamefont {Ando}, \citenamefont {Fowler},\ and\ \citenamefont {Stern}}]{Ando82}%
  \BibitemOpen
  \bibfield  {author} {\bibinfo {author} {\bibfnamefont {T.}~\bibnamefont {Ando}}, \bibinfo {author} {\bibfnamefont {A.~B.}\ \bibnamefont {Fowler}},\ and\ \bibinfo {author} {\bibfnamefont {F.}~\bibnamefont {Stern}},\ }\bibfield  {title} {\bibinfo {title} {Electronic properties of two-dimensional systems},\ }\href {https://doi.org/10.1103/RevModPhys.54.437} {\bibfield  {journal} {\bibinfo  {journal} {Rev. Mod. Phys.}\ }\textbf {\bibinfo {volume} {54}},\ \bibinfo {pages} {437} (\bibinfo {year} {1982})}\BibitemShut {NoStop}%
\bibitem [{\citenamefont {John}\ \emph {et~al.}(2004)\citenamefont {John}, \citenamefont {Castro},\ and\ \citenamefont {Pulfrey}}]{John04}%
  \BibitemOpen
  \bibfield  {author} {\bibinfo {author} {\bibfnamefont {D.~L.}\ \bibnamefont {John}}, \bibinfo {author} {\bibfnamefont {L.~C.}\ \bibnamefont {Castro}},\ and\ \bibinfo {author} {\bibfnamefont {D.~L.}\ \bibnamefont {Pulfrey}},\ }\bibfield  {title} {\bibinfo {title} {{Quantum capacitance in nanoscale device modeling}},\ }\href {https://doi.org/10.1063/1.1803614} {\bibfield  {journal} {\bibinfo  {journal} {Journal of Applied Physics}\ }\textbf {\bibinfo {volume} {96}},\ \bibinfo {pages} {5180} (\bibinfo {year} {2004})}\BibitemShut {NoStop}%
\bibitem [{\citenamefont {Xia}\ \emph {et~al.}(2009)\citenamefont {Xia}, \citenamefont {Chen}, \citenamefont {Li},\ and\ \citenamefont {Tao}}]{Xia09}%
  \BibitemOpen
  \bibfield  {author} {\bibinfo {author} {\bibfnamefont {J.}~\bibnamefont {Xia}}, \bibinfo {author} {\bibfnamefont {F.}~\bibnamefont {Chen}}, \bibinfo {author} {\bibfnamefont {J.}~\bibnamefont {Li}},\ and\ \bibinfo {author} {\bibfnamefont {N.}~\bibnamefont {Tao}},\ }\bibfield  {title} {\bibinfo {title} {Measurement of the quantum capacitance of graphene},\ }\href {https://doi.org/10.1038/nnano.2009.177} {\bibfield  {journal} {\bibinfo  {journal} {Nature Nanotechnology}\ }\textbf {\bibinfo {volume} {4}},\ \bibinfo {pages} {505} (\bibinfo {year} {2009})}\BibitemShut {NoStop}%
\bibitem [{\citenamefont {Mišković}\ and\ \citenamefont {Upadhyaya}(2010)}]{Miskovic010}%
  \BibitemOpen
  \bibfield  {author} {\bibinfo {author} {\bibfnamefont {Z.~L.}\ \bibnamefont {Mišković}}\ and\ \bibinfo {author} {\bibfnamefont {N.}~\bibnamefont {Upadhyaya}},\ }\bibfield  {title} {\bibinfo {title} {Modeling electrolytically top-gated graphene},\ }\href {https://doi.org/10.1007/s11671-009-9515-3} {\bibfield  {journal} {\bibinfo  {journal} {Nanoscale Res Lett.}\ }\textbf {\bibinfo {volume} {7}},\ \bibinfo {pages} {505} (\bibinfo {year} {2010})}\BibitemShut {NoStop}%
\bibitem [{\citenamefont {Christen}\ and\ \citenamefont {B\"uttiker}(1996)}]{Buttiker96}%
  \BibitemOpen
  \bibfield  {author} {\bibinfo {author} {\bibfnamefont {T.}~\bibnamefont {Christen}}\ and\ \bibinfo {author} {\bibfnamefont {M.}~\bibnamefont {B\"uttiker}},\ }\bibfield  {title} {\bibinfo {title} {Low frequency admittance of a quantum point contact},\ }\href {https://doi.org/10.1103/PhysRevLett.77.143} {\bibfield  {journal} {\bibinfo  {journal} {Phys. Rev. Lett.}\ }\textbf {\bibinfo {volume} {77}},\ \bibinfo {pages} {143} (\bibinfo {year} {1996})}\BibitemShut {NoStop}%
\bibitem [{\citenamefont {Buttiker}(1993)}]{Buttiker93}%
  \BibitemOpen
  \bibfield  {author} {\bibinfo {author} {\bibfnamefont {M.}~\bibnamefont {Buttiker}},\ }\bibfield  {title} {\bibinfo {title} {Capacitance, admittance, and rectification properties of small conductors},\ }\href {https://doi.org/10.1088/0953-8984/5/50/017} {\bibfield  {journal} {\bibinfo  {journal} {Journal of Physics: Condensed Matter}\ }\textbf {\bibinfo {volume} {5}},\ \bibinfo {pages} {9361} (\bibinfo {year} {1993})}\BibitemShut {NoStop}%
\bibitem [{\citenamefont {Wang}\ \emph {et~al.}(1998)\citenamefont {Wang}, \citenamefont {Guo}, \citenamefont {Mozos}, \citenamefont {Wan}, \citenamefont {Taraschi},\ and\ \citenamefont {Zheng}}]{Wang98}%
  \BibitemOpen
  \bibfield  {author} {\bibinfo {author} {\bibfnamefont {J.}~\bibnamefont {Wang}}, \bibinfo {author} {\bibfnamefont {H.}~\bibnamefont {Guo}}, \bibinfo {author} {\bibfnamefont {J.-L.}\ \bibnamefont {Mozos}}, \bibinfo {author} {\bibfnamefont {C.~C.}\ \bibnamefont {Wan}}, \bibinfo {author} {\bibfnamefont {G.}~\bibnamefont {Taraschi}},\ and\ \bibinfo {author} {\bibfnamefont {Q.}~\bibnamefont {Zheng}},\ }\bibfield  {title} {\bibinfo {title} {Capacitance of atomic junctions},\ }\href {https://doi.org/10.1103/PhysRevLett.80.4277} {\bibfield  {journal} {\bibinfo  {journal} {Phys. Rev. Lett.}\ }\textbf {\bibinfo {volume} {80}},\ \bibinfo {pages} {4277} (\bibinfo {year} {1998})}\BibitemShut {NoStop}%
\bibitem [{\citenamefont {Hou}\ \emph {et~al.}(2001)\citenamefont {Hou}, \citenamefont {Wang}, \citenamefont {Yang}, \citenamefont {Wang}, \citenamefont {Wang}, \citenamefont {Zhu},\ and\ \citenamefont {Xiao}}]{Hou01}%
  \BibitemOpen
  \bibfield  {author} {\bibinfo {author} {\bibfnamefont {J.~G.}\ \bibnamefont {Hou}}, \bibinfo {author} {\bibfnamefont {B.}~\bibnamefont {Wang}}, \bibinfo {author} {\bibfnamefont {J.}~\bibnamefont {Yang}}, \bibinfo {author} {\bibfnamefont {X.~R.}\ \bibnamefont {Wang}}, \bibinfo {author} {\bibfnamefont {H.~Q.}\ \bibnamefont {Wang}}, \bibinfo {author} {\bibfnamefont {Q.}~\bibnamefont {Zhu}},\ and\ \bibinfo {author} {\bibfnamefont {X.}~\bibnamefont {Xiao}},\ }\bibfield  {title} {\bibinfo {title} {Nonclassical behavior in the capacitance of a nanojunction},\ }\href {https://doi.org/10.1103/PhysRevLett.86.5321} {\bibfield  {journal} {\bibinfo  {journal} {Phys. Rev. Lett.}\ }\textbf {\bibinfo {volume} {86}},\ \bibinfo {pages} {5321} (\bibinfo {year} {2001})}\BibitemShut {NoStop}%
\bibitem [{\citenamefont {Pan}\ \emph {et~al.}(1999)\citenamefont {Pan}, \citenamefont {Hudson},\ and\ \citenamefont {Davis}}]{pan1999he}%
  \BibitemOpen
  \bibfield  {author} {\bibinfo {author} {\bibfnamefont {S.}~\bibnamefont {Pan}}, \bibinfo {author} {\bibfnamefont {E.~W.}\ \bibnamefont {Hudson}},\ and\ \bibinfo {author} {\bibfnamefont {J.}~\bibnamefont {Davis}},\ }\bibfield  {title} {\bibinfo {title} {He 3 refrigerator based very low temperature scanning tunneling microscope},\ }\href@noop {} {\bibfield  {journal} {\bibinfo  {journal} {Review of scientific instruments}\ }\textbf {\bibinfo {volume} {70}},\ \bibinfo {pages} {1459} (\bibinfo {year} {1999})}\BibitemShut {NoStop}%
\bibitem [{\citenamefont {M.Crowley}(2008)}]{Crowley08}%
  \BibitemOpen
  \bibfield  {author} {\bibinfo {author} {\bibfnamefont {J.}~\bibnamefont {M.Crowley}},\ }\bibfield  {title} {\bibinfo {title} {{Simple Expressions for Force and Capacitance for a Conductive Sphere near a Conductive Wall}},\ }in\ \href {https://api.semanticscholar.org/CorpusID:201801924} {\emph {\bibinfo {booktitle} {Proc. ESA Annual Meeting on Electrostatics}}},\ \bibinfo {series and number} {Paper D1}\ (\bibinfo {year} {2008})\BibitemShut {NoStop}%
\bibitem [{\citenamefont {Kurokawa}\ and\ \citenamefont {Sakai}(1998)}]{Kurokawa98}%
  \BibitemOpen
  \bibfield  {author} {\bibinfo {author} {\bibfnamefont {S.}~\bibnamefont {Kurokawa}}\ and\ \bibinfo {author} {\bibfnamefont {A.}~\bibnamefont {Sakai}},\ }\bibfield  {title} {\bibinfo {title} {{Gap dependence of the tip-sample capacitance}},\ }\href {https://doi.org/10.1063/1.367985} {\bibfield  {journal} {\bibinfo  {journal} {Journal of Applied Physics}\ }\textbf {\bibinfo {volume} {83}},\ \bibinfo {pages} {7416} (\bibinfo {year} {1998})}\BibitemShut {NoStop}%
\bibitem [{\citenamefont {Roth}\ \emph {et~al.}(2017)\citenamefont {Roth}, \citenamefont {Bruckner}, \citenamefont {Moro}, \citenamefont {Gruber}, \citenamefont {Goebl}, \citenamefont {Juaristi}, \citenamefont {Alducin}, \citenamefont {Steinberger}, \citenamefont {Duchoslav}, \citenamefont {Primetzhofer} \emph {et~al.}}]{roth2017electronic}%
  \BibitemOpen
  \bibfield  {author} {\bibinfo {author} {\bibfnamefont {D.}~\bibnamefont {Roth}}, \bibinfo {author} {\bibfnamefont {B.}~\bibnamefont {Bruckner}}, \bibinfo {author} {\bibfnamefont {M.}~\bibnamefont {Moro}}, \bibinfo {author} {\bibfnamefont {S.}~\bibnamefont {Gruber}}, \bibinfo {author} {\bibfnamefont {D.}~\bibnamefont {Goebl}}, \bibinfo {author} {\bibfnamefont {J.}~\bibnamefont {Juaristi}}, \bibinfo {author} {\bibfnamefont {M.}~\bibnamefont {Alducin}}, \bibinfo {author} {\bibfnamefont {R.}~\bibnamefont {Steinberger}}, \bibinfo {author} {\bibfnamefont {J.}~\bibnamefont {Duchoslav}}, \bibinfo {author} {\bibfnamefont {D.}~\bibnamefont {Primetzhofer}}, \emph {et~al.},\ }\bibfield  {title} {\bibinfo {title} {Electronic stopping of slow protons in transition and rare earth metals: breakdown of the free electron gas concept},\ }\href@noop {} {\bibfield  {journal} {\bibinfo  {journal} {Physical Review Letters}\ }\textbf {\bibinfo {volume} {118}},\ \bibinfo {pages} {103401} (\bibinfo {year} {2017})}\BibitemShut
  {NoStop}%
\bibitem [{\citenamefont {{de Ara}}\ \emph {et~al.}(2022)\citenamefont {{de Ara}}, \citenamefont {Sabater}, \citenamefont {Borja-Espinosa}, \citenamefont {Ferrer-Alcaraz}, \citenamefont {Baciu}, \citenamefont {Guijarro},\ and\ \citenamefont {Untiedt}}]{DeAra22}%
  \BibitemOpen
  \bibfield  {author} {\bibinfo {author} {\bibfnamefont {T.}~\bibnamefont {{de Ara}}}, \bibinfo {author} {\bibfnamefont {C.}~\bibnamefont {Sabater}}, \bibinfo {author} {\bibfnamefont {C.}~\bibnamefont {Borja-Espinosa}}, \bibinfo {author} {\bibfnamefont {P.}~\bibnamefont {Ferrer-Alcaraz}}, \bibinfo {author} {\bibfnamefont {B.~C.}\ \bibnamefont {Baciu}}, \bibinfo {author} {\bibfnamefont {A.}~\bibnamefont {Guijarro}},\ and\ \bibinfo {author} {\bibfnamefont {C.}~\bibnamefont {Untiedt}},\ }\bibfield  {title} {\bibinfo {title} {Signature of adsorbed solvents for molecular electronics revealed via scanning tunneling microscopy},\ }\href {https://doi.org/https://doi.org/10.1016/j.matchemphys.2022.126645} {\bibfield  {journal} {\bibinfo  {journal} {Materials Chemistry and Physics}\ }\textbf {\bibinfo {volume} {291}},\ \bibinfo {pages} {126645} (\bibinfo {year} {2022})}\BibitemShut {NoStop}%
\bibitem [{\citenamefont {Martinez-Garcia}\ \emph {et~al.}(2023)\citenamefont {Martinez-Garcia}, \citenamefont {de~Ara}, \citenamefont {Pastor-Amat}, \citenamefont {Untiedt}, \citenamefont {Lombardi}, \citenamefont {Dednam},\ and\ \citenamefont {Sabater}}]{Martinez23}%
  \BibitemOpen
  \bibfield  {author} {\bibinfo {author} {\bibfnamefont {A.}~\bibnamefont {Martinez-Garcia}}, \bibinfo {author} {\bibfnamefont {T.}~\bibnamefont {de~Ara}}, \bibinfo {author} {\bibfnamefont {L.}~\bibnamefont {Pastor-Amat}}, \bibinfo {author} {\bibfnamefont {C.}~\bibnamefont {Untiedt}}, \bibinfo {author} {\bibfnamefont {E.~B.}\ \bibnamefont {Lombardi}}, \bibinfo {author} {\bibfnamefont {W.}~\bibnamefont {Dednam}},\ and\ \bibinfo {author} {\bibfnamefont {C.}~\bibnamefont {Sabater}},\ }\bibfield  {title} {\bibinfo {title} {Unraveling the interplay between quantum transport and geometrical conformations in monocyclic hydrocarbons’ molecular junctions},\ }\href {https://doi.org/10.1021/acs.jpcc.3c05393} {\bibfield  {journal} {\bibinfo  {journal} {The Journal of Physical Chemistry C}\ }\textbf {\bibinfo {volume} {127}},\ \bibinfo {pages} {23303} (\bibinfo {year} {2023})},\ \Eprint {https://arxiv.org/abs/https://doi.org/10.1021/acs.jpcc.3c05393} {https://doi.org/10.1021/acs.jpcc.3c05393} \BibitemShut {NoStop}%
\end{thebibliography}%

\end{document}